\documentclass[10pt,conference]{IEEEtran}
\usepackage{epsfig, epsf, amsthm, amsmath, amssymb, amsfonts, subfigure,color}

\usepackage{cite}
\usepackage{graphicx}
\usepackage{textcomp}
\usepackage{xcolor}
\usepackage{dblfloatfix} 
\usepackage{enumitem}
\usepackage{url}
\usepackage{bbm}
\usepackage{algorithm,algorithmic,xpatch}


\def\l{\left}
\def\r{\right}
\def\({\l(}
\def\){\r)}
\def\[{\l[}
\def\]{\r]}
\def\T{\intercal}

\IEEEoverridecommandlockouts

\begin{document}
	
	\title{  Remote UAV Online Path Planning via \\ Neural Network Based Opportunistic Control
	\\
		\thanks{}
	}
	
	\author{\IEEEauthorblockN{Hamid Shiri, Jihong Park, and Mehdi Bennis}
		\IEEEauthorblockA{
			\textit{University of Oulu, Finland}\\
			$\{$hamid.shiri, jihong.park, mehdi.bennis$\}$@oulu.fi}
	}
	
	\maketitle
	
\begin{abstract}
	This letter proposes a neural network (NN) aided remote unmanned aerial vehicle (UAV) online control algorithm, coined \emph{{oHJB}}. By downloading a UAV's state, a base station (BS) trains an HJB NN that solves the Hamilton-Jacobi-Bellman equation (HJB) in real time, yielding the optimal control action. Initially, the BS uploads this control action to the UAV. If the HJB NN is sufficiently trained and the UAV is far away, the BS uploads the HJB NN model, enabling to locally carry out control decisions even when the connection is lost. Simulations corroborate the effectiveness of {oHJB} in reducing the UAV's travel time and energy by utilizing the trade-off between uploading delays and control robustness in poor channel conditions.
\end{abstract}
	
	\begin{IEEEkeywords}
		Remote UAV control, communication-efficient online path planning, machine learning.
	\end{IEEEkeywords}
	
\section{Introduction}

Controlling unmanned aerial vehicles (UAVs) over wireless links is one key use case in 5G and beyond, facilitating new mission-critical applications ranging from emergency wireless networks to rescue missions~\cite{Fan2018Lett, b40,Ackerman:18}. To enable this, there are two fundamental challenges. Firstly, UAV states (e.g., positions and velocities) are time-varying and perturbed by wind dynamics, which requires a real-time control algorithm with low computational complexity~\cite{b46}. Secondly, each UAV's state sensor and actuator are not co-located with its controller, as Fig.~1 illustrates. Due to wireless channel fluctuations, connecting them with low latency without compromising the control optimality is a daunting task. Spurred by these issues, in this letter, we propose a communication-efficient remote UAV online control algorithm via machine learning (ML).

\vspace{4pt}\noindent\textbf{ML Aided Real-Time Control.} \quad
For a given continuous-time state dynamics, an optimal control action can be obtained by solving the \emph{Hamilton-Jabobi-Bellman equation (HJB)} \cite{Aguilar2014, b2}, i.e., a partial differential equation $\mathsf{H}=0$. Unfortunately, solving the HJB equation is challenging when the states are larger than 1-dimensional~\cite{b50,b3}. Alternatively, the HJB solution can be approximated by a neural network (NN) that minimizes the norm of $ \mathsf{H} $, i.e., $|\mathsf{H}|$ \cite{b2, shiri2019massive}, referred to as the \emph{HJB NN} that outputs the optimal control action of its input state.

\vspace{4pt}\noindent\textbf{Uploading Actions} (\textsf{aHJB}) \textbf{vs. Models} (\textsf{mHJB})\textbf{.} \quad
As shown in Fig.~1, each state of a UAV is downloaded by its remote controller, i.e., base station (BS), running the HJB NN. To upload the HJB NN results to the UAV, we propose two baseline choices at the BS. One way is to upload the optimal control action (i.e., the NN output), denoted as {\textsf{aHJB}}. This is however not preferable under poor channel conditions, in which the computed actions are easily outdated. The other way is to upload the current HJB NN model, denoted as {\textsf{mHJB}}. Once the HJB NN is uploaded, the UAV can locally carry out its control decisions even when the connection is lost. Since NN model sizes are often larger than action command sizes, there is a \emph{trade-off between uploading delays and control robustness} against poor channel conditions. 

\vspace{4pt}\noindent\textbf{Opportunistic HJB NN Control} (\textsf{oHJB})\textbf{.} \quad
To resolve the aforementioned trade-off, we propose {\textsf{oHJB}} wherein the BS switches from \textsf{aHJB} to \textsf{mHJB} if the HJB NN is sufficiently trained and the UAV is far away from the BS. To examine these opportunities, the former is quantified by the number of downloaded states, and the latter is determined by the average delay during $N$ latest control time slots.

We examine the effectiveness of \textsf{oHJB}, in terms of UAV travel time and motion energy consumption until reaching a destination from a source. Additionally, we study the impact of uplink transmission power control of the BS based on the downlink delays. Compared to \textsf{aHJB} and \textsf{mHJB}, numerical results corroborate that \textsf{oHJB} achieves faster travel time with less energy consumption, even without power control.

\begin{figure}[t] \label{fig:01} 
	\centering
	\includegraphics[width=.87\columnwidth, trim={0 .5cm 0 0}]{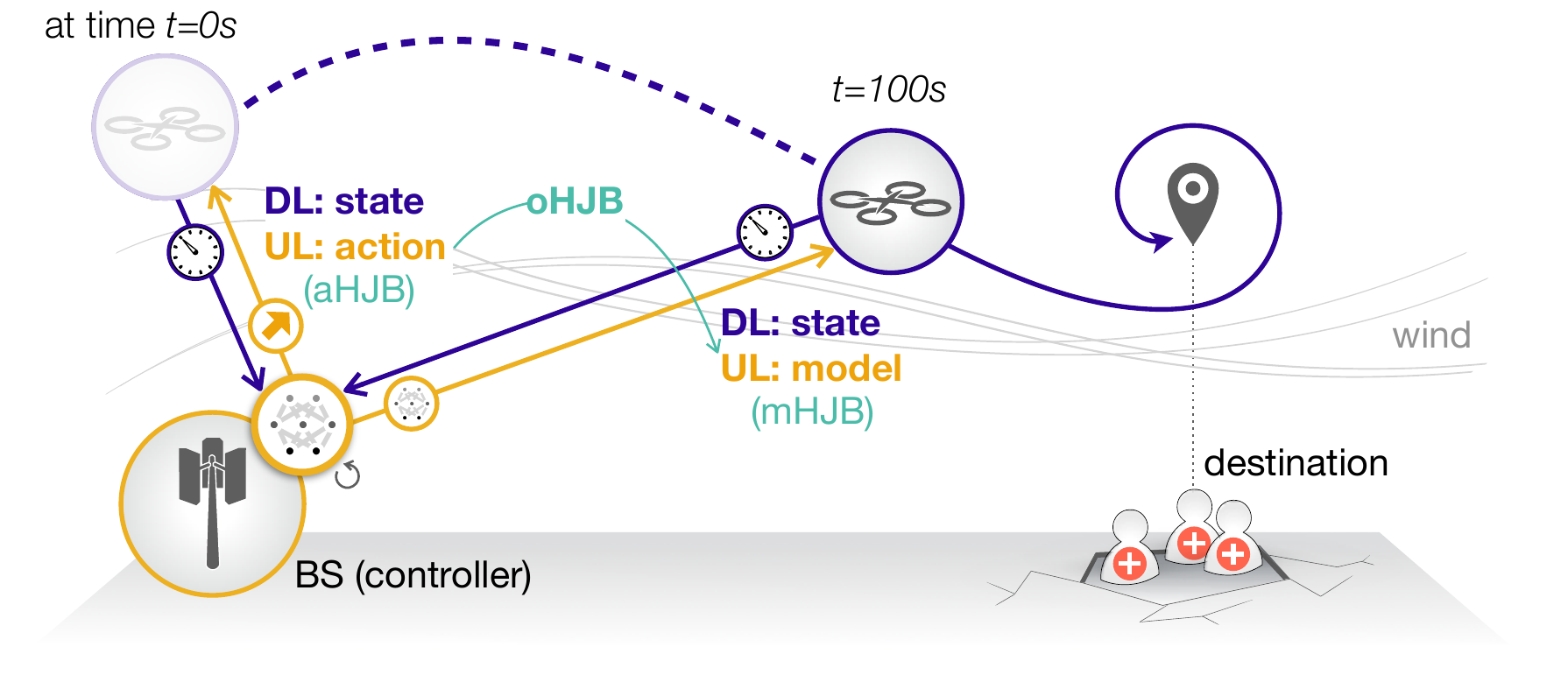}
	\caption{ An illustration of the \emph{opportunistic HJB NN control (\textsf{oHJB})} of a UAV, switching from action-uploading (\textsf{aHJB}) at time $t=0$s to model-uploading (\textsf{mHJB}) operations at $t=100$s.}
	\vspace{-2.0em}
\end{figure}

	\begin{figure*}[b!]
		\hrulefill \small
		\setcounter{equation}{3}
		\footnotesize\begin{align} 
		\label{Eq:HJB}
		\hspace{-43pt}\hspace{5pt} \mathsf{H}\big(\psi (t);s(t) \big)& = \partial_{t} \psi (t) + \inf_{a(t)} \bigg\{ \big[ A s(t) + B (a(t) + c_0 v_o ) \big]^{\T}  \nabla \psi(t)
		+  \frac{1}{2}\textup{tr}\!\(G G^{\T} \nabla^2 \psi\!(t)\) + \phi_L\!\(s(t)\) + c_3 {\left\| a(t) \right\|^{2}  }    \bigg\}\\
		\label{Eq:HJBmodified}
		&= \partial_{t} \psi (t)  +   \[ A s(t)   -  \frac{1}{4 c_3} B B^{\T} \nabla \psi(t)  +  c_0 B v_o  \]^{\T}  \nabla \psi(t)  
		+ \frac{1}{2}\textup{tr}\!\(G G^{\T} \nabla^2 \psi(t)\) +   \phi_L\!\(s(t)\)  
		\end{align} \normalsize
	\end{figure*}
	\setcounter{equation}{7}


\vspace{4pt}\noindent \textbf{Related works}.\quad
Offline UAV deployment and path planning have been studied in \cite{b40,Shashi2007offlineUAV}. For online path planning, autonomous UAV control scenarios have been investigated in \cite{Rob2018UAVGA,RL2018UAVhuy,shiri2019massive}. Among them, single autonomous control methods have been proposed in \cite{Rob2018UAVGA,RL2018UAVhuy}, using a genetic algorithm \cite{Rob2018UAVGA} and Q-learning with the PID controller \cite{RL2018UAVhuy}, respectively. On the other hand, a control scheme for massive autonomous UAVs has been proposed in \cite{shiri2019massive}, by leveraging a NN based mean-field game theoretic algorithm. However, these works ignore communication efficiency, or focus only on short-range inter-UAV communications. Long-range BS-to-UAV communications have been studied in \cite{Comm19UAVmardani, b48,b3}, in the context of offline UAV deployment and path planning. This letter aims to bridge this gap and propose a communication-efficient remote UAV online path planning method using an NN based algorithm.



\section{Online Learning Aided Optimal Actions} \label{SE:02}
In this section, we formulate an optimization problem of controlling a single UAV, followed by its solution using HJB NN learning. This problem is solved by a BS locally running the HJB NN, and the communications between the BS and UAV will be elaborated in Sec. III.

	\subsection{Control Problem Formulation}
	We consider a single UAV and a BS whose altitudes are fixed as $h>0$ and $0$, respectively, and focus on the movement of the UAV in a two-dimensional plane. Initially, both UAV and BS are located at a source $r_s\in\mathbb{R}^2$, and the UAV is controlled to reach a destination located at the origin. Let $s(t)= [r(t)^{\T},v(t)^{\T}]^{\T}\in\mathbb{R}^4$ denote the UAV state describing the location $r(t)\in\mathbb{R}^2$ and velocity $v(t)\in\mathbb{R}^2$ of the UAV at time $t\geq 0$. The BS aims to minimize the source-to-destination travel time and motion energy of the UAV, by adjusting the acceleration $a(t)\in\mathbb{R}^2$, i.e., control action. This is cast by minimizing the average cost $\psi(t)$, given as:

\setcounter{equation}{0}
\vspace{-10pt}\small
	\begin{align} \label{Eq:LRA}
	\hspace{-5pt} \psi(t) \!=\!  \mathsf{E} \bigg[\! \int_{t}^{T}\!\!  \bigg(\!\! 
	\underbrace{\frac{v(\tau) r(\tau)}{\left \| r(\tau) \right \|} +  c_1 {\left\| r(\tau) \right\|^{2}}  \!+ {c_2 {\left\| v(\tau) \right\|^{2}}} }_{\phi_L\(s(\tau)\)}
	\!+ c_3 {\left\| a(\tau) \right\|^{2}  }
	\!\bigg) \textup{d}\tau \!\bigg],  
	\end{align}\normalsize 
	where the average is taken across all possible actions for $\tau\!\in\![t,T]$, and the terms $ c_1 $, $ c_2 $, and $ c_3 $ are positive constants. Inside the integration, the first and second terms are intended to minimize the travel time, by increasing the velocity towards the destination, i.e., ${v(\tau)r(\tau)}/{|| r(\tau) ||} $, and by decreasing the remaining travel distance, i.e., $||r(t)||^2$, respectively. The third and last terms aim to minimize the motion energy consumed by the kinetic energy, i.e., $|| v(\tau)||^2$, and the controlled actuation, i.e., $||a(\tau)||^2$, respectively.

	In addition, following \cite{b49,b3,shiri2019massive}, we consider wind dynamics characterized by the average wind velocity $v_o\in\mathbb{R}^2$ and random perturbations following a Wiener process $W(t)$ with the covariance matrix $V_o\in\mathbb{R}^{2\times 2}$. Consequently, at time $t$, the UAV control problem subject to the UAV state dynamics, is formulated as:
	
	\vspace{-10pt}
	\small
	\begin{align}
	&\psi^*\!(t) = \min_{a(t)}\; \psi(t),  \label{Eq:psi} \\
	\text{s.t.}\quad &\text{d} s(t)= \left( A s(t) + B (a(t) + c_0 v_o ) \right) \textup{d}t + G \text{d} W(t), \label{Eq:StateDyn}
	\end{align}\normalsize
	where
	$A\!=\!\(\begin{smallmatrix}O & I\\O & -c_0 I\end{smallmatrix}\)$, $B\!=\!\(\begin{smallmatrix}O \\ 
	I\end{smallmatrix}\)$, $G\!=\!\(\begin{smallmatrix}O \\ V_o\end{smallmatrix}\)$, and $ c_0 $ is a positive constant. The terms $I$ and $ O $ denote the two-dimensional identity matrix and zero matrix, respectively. 
	The minimum cost $\psi^*\!(t)$ is called as the \emph{value function} of the optimal control, and is approximated via the HJB NN as detailed in the following subsection.

\begin{figure*}[b]
\setcounter{equation}{10}\hrule \small
\begin{align}
L(t) = \begin{cases}
\underbrace{30.9+(22.25-0.5\log_{10}h)\log_{10}d(t)+20\log_{10}f_c}_{L_\text{LOS}}
&\qquad \text{if LOS}\\
\max\{L_\text{LOS},\; 32.4+(43.2-7.6\log_{10}h)\log_{10}d(t)+20\log_{10}f_c\}
&\qquad \text{otherwise,} \end{cases} \label{pathloss_1}
\end{align}
where $d(t)=\Arrowvert r(t)-r_s \Arrowvert$ and $f_c$ is the carrier frequency. The LOS conditions occur with probability $p_\text{LOS}$ given as:

\vspace{-10pt}
\begin{align}
p_\text{LOS}\!\(d(t)\) = \begin{cases} {d_o}/{\sqrt{d(t)^2-h^2}}+\exp\left\{\left( {-\sqrt{d(t)^2-h^2}}/{p_o}\right ) \!\left( 1-{d_o}/{\sqrt{d(t)^2-h^2}}  \right) \right\} & \text{if }\sqrt{d(t)^2-h^2}    >  d_o
\\  1  &\text{otherwise},  \label{pathloss_2}
\end{cases}
\end{align}
where ${d_o\!=\!\max\{294.05\log_{10}h\!-\!432.94,18\}}$ and $p_o=233.98\log_{10}h-0.95$.
\end{figure*}

	\subsection{Optimal Action Calculation via Online HJB NN Learning} \label{Sec:HJBcontrol}
	According to \cite{b12}, the value function $\psi^*\!(t)$ satisfies the HJB $\mathsf{H}\big(\psi^*\!(t);s(t) \big)\!=\!0$ having a unique solution, where the Hamiltonian function $\mathsf{H}\big(\psi^*\!(t);s(t) \big)$ is given in \eqref{Eq:HJBmodified} at the bottom of this page (see the derivation details in~\cite{b12}). Since $\mathsf{H}\big(\psi^*\!(t);s(t) \big)$ is convex with respect to $a(t)$, the optimal action $a^*(t)$ is obtained by the first-order condition, yielding:

	\setcounter{equation}{5}	
	\vspace{-10pt}\small\begin{align}  \label{Eq:opt_act}
	a^*(t) \!=\! - \frac{1}{2c_3}B^\T\nabla\psi^*\!(t), 
	\end{align}\normalsize
	where the gradient $\nabla$ is taken with respect to $s(t)$. However, solving the HJB incurs huge computational complexity particularly for multi-dimensional states~\cite{b50}, which is ill-suited for real-time UAV control.

	To resolve this issue, following \cite{shiri2019massive}, we approximate $ \psi^*\!(t) $ as $ \hat{\psi}(t) $, and recast \eqref{Eq:opt_act} by $a^*(t)\approx- B^\T \nabla \hat{\psi}\!(t)/(2 c_3)$. For obtaining $ \hat{\psi}(t) $, we utilize the HJB NN, a single-layer NN with $M$ weights, resulting in:
	
	\vspace{-10pt}\small
	\begin{align} \label{Eq:07}
	\hat{\psi}(t) := w(t)^\T \sigma\!\(s(t)\),
	\end{align} \normalsize
	where $ w(t) \in\mathbb{R}^{M}$ is a weight vector and $ \sigma\!\(s(t)\) \in\mathbb{R}^{M}$ is an activation vector. The HJB NN is trained by adjusting $ w(t) $ for each observation $ s(t) $, so as to minimize the following loss function $\mathcal{L}(t)$:
	
	\vspace{-10pt}\small\begin{align} \label{Eq:HJBlearncost}
	\hspace{-6pt} \mathcal{L}(t) = \underbrace{\frac{1}{2} \l| {\mathsf{H}}\big(\hat{\psi}(t);s(t)\big) \r|^{2}}_{\ell(t)} +\; c_{\Omega}  \underbrace{\max\l\{0, s(t)^\T\frac{\text{d}s(t)}{\text{d}t}\r\} }_{\Omega(t)},
	\end{align}\normalsize \vskip -5pt
	\noindent where $c_{\Omega}$ is a positive constant. Here, minimizing $\ell(t)$ approximates solving the HJB, i.e., $\hat{\psi}(t)\approx \psi^*\!(t)$. The regularizer $\Omega(t)$ is intended to stop the movement when reaching the destination, i.e., $s(T)=0$. While learning the HJB NN, $\mathcal{L}(t)$ is minimized by the normalized gradient descent algorithm (NGD) \cite{b51} updating the weights as:
	
	\vspace{-10pt}\small\begin{align} \label{Eq:08}
	\hspace{-5pt} w(t) \!=\! w_{\psi}(t') \!-\! \mu \text{sign}\(\nabla_{\!w} \ell(t)\) \!-\! c_{\Omega}\nabla_{\!w} \Omega(t), 
	\end{align}\normalsize 
	where $\mu>0$ is the step size, $\text{sign}(x) = x/\| x\|$, and $ t' $ indicates the time of the previous update.

\section{Remote Control Over Wireless Links}
In this section, we describe the downlink and uplink communication latencies between the BS and the UAV. Based on this, we introduce two remote UAV control baselines, \textsf{aHJB} and \textsf{mHJB}, and finally propose \textsf{oHJB}.
	
\subsection{Communication Protocol and Delay Model}
The $k$-th optimal UAV control entails both downlink and uplink delays. Specifically, $D_{\text{DL},k} = t_{\text{DL},k}-t_{o,k}$ is the downlink delay between the BS's state downloading completion time $t_{\text{DL},k}$ and its downloading starting time $t_{o,k}$, at which the UAV state is measured. The uplink delay $D_{\text{UL},k}= t_{\text{UL},k}-t_{\text{DL},k}$ is the time difference between the BS's uploading completion $t_{\text{UL},k}$ and $t_{\text{DL},k}$. Assuming the HJB NN weight updating time is negligible, the end-to-end (E2E) delay of the $k$-th control is $D_{\text{DL},k} + D_{\text{UL},k}= t_{\text{UL},k}-t_{o,k}$. To make this E2E delay bounded, the UAV waits for the uploading information until a deadline $D_\text{th}$ from $t_{o,k}$, and then transmits the next state measured at $t_{o,k+1}$. Let the subscript $i\in\{\text{DL}, \text{UL}\}$ identify the downlink and uplink, and $D_{i,k}$ is detailed next.

In order to ensure short control intervals, $D_\text{th}$ is set as a sufficiently small value. In this case, the path loss $L(t)$ can be considered as a constant during the $k$-th control, i.e., $L(t)= L(t_{o,k})$ $\forall t\in [t_{o,k},t_{o,k+1})$, while $L(t)$ is assumed to be independent and identically distributed (i.i.d.) across different $k$'s. Consequently, the downlink or uplink delay $D_{i,k}$ is:

\vspace{-10pt}\small\begin{align}
D_{i,k}=\frac{b_i}{W_i \log_2\(1 + P_i \!\cdot\! 10^{-L(t_{o,k})/10}\! / (N_i W_i) \)}, \label{Eq:channel_rate}
\end{align}\normalsize
where $W_i$ is the frequency bandwidth, $P_i$ is the transmission power, and $ N_i $ is the noise spectral density. 

The distribution of the path loss $L(t)$ follows the 3GPP urban-micro UAV channel model \cite{3GPP17}, which is applicable for the UAV altitude $ 22.5\text{m}  \leq h \leq 300\text{m} $. This channel model describes $L(t)$ in \eqref{pathloss_1} with line-of-sight (LOS) probability $p_\text{LOS}\!\(d(t)\)$ in \eqref{pathloss_2}, given at the bottom of this page.

The downlink transmission power $P_{\text{DL}}$ of the UAV is set as a constant. On the other hand, the uplink transmission power $P_{\text{UL}}$ of the BS is by default set as $P_{\text{UL},o}$, and increased to $P_{\text{UL},\text{max}}$ if the preceding downlink delay $D_{\text{DL},k}$ is too long and $P_{\text{UL},o}$ is insufficient, given as:

\setcounter{equation}{12}
\vspace{-10pt}\small
	\begin{align} \label{Eq:power_control}
	P_{\text{UL}} = \begin{cases}
	P_{\text{UL},o} & D_{\text{DL},k} \leq D_{\text{th}} - \hat{D}_{\text{UL},k}\\ 
	P_{\text{UL},\text{max}} & \text{otherwise},
	\end{cases}
\end{align} \normalsize 
where {\small$\hat{D}_{\text{UL},k}=b_\text{UL}/[W_{\text{UL}}\log_2 (1 + P_{\text{UL},o} 10^{-L(t_{o,k}/10)}\!/(N_i W_i))]$\normalsize}. This term can be evaluated based on $P_{\text{UL},o}$ under the fully correlated downlink and uplink channels, i.e., $L(t)= L(t_{o,k})$ $\forall t\in [t_{o,k},t_{o,k+1})$.

\begin{algorithm}[t] \label{alg:01}
		\caption{\textsf{aHJB} control}
		\begin{algorithmic}[1]
			\STATE \textbf{Initialization:} $ s(0) $, $ w(t_{\text{DL},0}) = 0 $.
			
			\WHILE {$  |r(t_{o,k})| > r_\text{th} $ and $ |v(t_{o,k})| > v_\text{th} $}
			\STATE UAV \textbf{sends} the state $ s(t_k) $ to the BS.
			\STATE BS \textbf{updates} the model parameters $w(t_{\text{DL},k})$ via \eqref{Eq:08}. 
			
			\STATE BS \textbf{sends} the action $ a^*(t_{\text{DL},k}) $ calculated via \eqref{Eq:opt_act} using transmit power calculated via \eqref{Eq:power_control}.
			
			\STATE UAV \textbf{takes} the most recently received action $ a^*(t_{\text{DL},k'}) $. 
			\ENDWHILE
		\end{algorithmic} 
	\end{algorithm} 
	\begin{algorithm}[t] \label{alg:02}
		\caption{\textsf{mHJB} control}
		\begin{algorithmic}[1]
			\STATE \textbf{Initialization:} $ s(0) $, $ w(t_{\text{DL},0}) = 0 $.
			
			\WHILE {$  |r(t_{o,k})| > r_\text{th} $ and $ |v(t_{o,k})| > v_\text{th} $}
			
			\STATE UAV \textbf{sends} the state $ s(t_k) $ to the BS.
			\STATE BS \textbf{updates} the model parameters $w(t_{\text{DL},k})$ via \eqref{Eq:08}. 
			\STATE BS \textbf{sends} the model parameters $ w(t_{\text{DL},k}) $  using transmit power calculated via \eqref{Eq:power_control}.
			
			\STATE UAV uses the most recently received model parameters $ w(t_{\text{DL},k'}) $ to calculate and \textbf{take} the action $ a^*(t_{\text{DL},k}) $. 
			\ENDWHILE 
		\end{algorithmic} 
	\end{algorithm}

\subsection{Baselines: aHJB and mHJB Control Algorithms}
The operations of \textsf{aHJB} are summarized in Algorithm 1. In the downlink phase, at the $k$-th control, the UAV state $s(t_{o,k})$ is received by the BS. In the HJB NN learning phase, $s(t_{o,k})$ is fed to the HJB NN, updating its previous weight vector $w(t_{\text{DL},k-1})$ to $w(t_{\text{DL},k})$, according to \eqref{Eq:08} with $t=t_{\text{DL},k}$ and $t'=t_{\text{DL},k-1}$. In the uplink phase, based on the HJB NN output $\hat{\psi}(t_{\text{DL,k}})$, the optimal action $ a^*(t_{\text{DL},k})$ in \eqref{Eq:opt_act} is calculated and transmitted to the UAV. The UAV continues to use the most recently received optimal action until the next action is received. The UAV control mission ends when the remaining distance and the velocity are small enough, i.e., $|r(t_{o,k})| \leq r_\text{th} $ and $ |v(t_{o,k})| \leq v_\text{th}$ for target thresholds $r_\text{th}> 0$ and $v_\text{th}>0$.

Next, the operations of \textsf{mHJB} control are summarized in Algorithm 2. In \textsf{mHJB}, the downlink and HJB NN learning phases follow the same procedures of \textsf{aHJB}. In the uplink phase, the current HJB NN model, i.e., the weight vector $w(t_{\text{DL},k})$, is transmitted to the UAV. By feeding the current state $s(t)$ to the uploaded HJB NN, the UAV can locally evaluate $a^*(t)$ in \eqref{Eq:opt_act} in real time, thereby carrying out the optimal control decisions regardless of the channel conditions. 

In the aforementioned operations, the downlink payload size is given as $b_s= 4  b$, where $4$ comes from the state dimension in Sec. II-A and $b$ is determined by the arithmetic precision. In the uplink, the payload sizes $b_{a}$ and $b_{m}$ of \textsf{aHJB} and \textsf{mHJB} are given as $b_a = 2  b$ and $b_m = M  b$, respectively, in which $2$ corresponds to the action dimension in Sec. II-A and $M$ comes from the weight vector size in Sec.~II-B.

\begin{algorithm}[t] \label{alg:03}
		\caption{\textsf{oHJB} control}
		\begin{algorithmic}[1]
			\STATE \textbf{Initialization:} $ s(0) $, $ w(t_{\text{DL},0}) = 0 $.
			
			\WHILE {$  |r(t_{o,k})| > r_\text{th} $ and $ |v(t_{o,k})| > v_\text{th} $}
			
			\STATE UAV \textbf{sends} the state $ s(t_k) $ to the BS.
			\STATE BS \textbf{updates} the model parameters $w(t_{\text{DL},k})$ via \eqref{Eq:08}.   
			
			\IF {$  \text{Dn}(t) \geq  \text{Dn}_{th} $ \OR $ \bar{D}_{\text{DL},k} \leq \alpha D_{\text{th}} $}
			
			\STATE BS \textbf{sends} the action $ a^*(t_{\text{DL},k}) $ calculated via \eqref{Eq:opt_act} with power calculated via \eqref{Eq:power_control}.
			\STATE UAV \textbf{takes} the most recently received action $ a^*(t_{\text{DL},k'}) $. 
			
			\ELSE 
			
			\STATE BS \textbf{sends} the model parameters $ w(t_{\text{DL},k}) $  with power calculated via \eqref{Eq:power_control}.
			\STATE UAV uses the most recently received model parameters $ w(t_{\text{DL},k'}) $ to calculate and \textbf{take} the action $ a^*(t_{\text{DL},k}) $. 
			
			\ENDIF
			\ENDWHILE
		\end{algorithmic}
	\end{algorithm}

\subsection{Proposed: oHJB Control Algorithm}	
There exists a trade-off between \textsf{aHJB} and \textsf{mHJB}. At the beginning, the HJB NN is not fully trained, and the NN outputs cannot accurately approximate the optimal actions. In this case, quickly completing uplink transmissions allows the BS to download more UAV states, thereby training the HJB NN more frequently. In the early phase of the UAV operations, \textsf{aHJB} is therefore preferable, compared to \textsf{mHJB} whose HJB NN model size is much larger than the action dimension (e.g., $M=54\gg 2$ in Sec. V). This strategy is also advocated given the small path loss for small $t$, when the UAV is close to the BS. On the contrary, the HJB NN is well trained for a sufficiently large $t$. By uploading this model, the UAV can locally carry out optimal control decisions, even when it becomes far away from the BS.

Motivated by this trade-off, we propose \textsf{oHJB} that follows \textsf{aHJB} at the beginning and switches to \textsf{mHJB} at a certain time, as summarized in Algorithm 3. The switching time is determined by satisfying the following two conditions: (i) the number of downloaded states $\text{Dn}(t)$ is larger than a threshold $\text{Dn}_\text{th}>0$; and (ii) the average downloading delay $\bar{D}_{\text{DL},k} =  \frac{1}{N}\sum_{m=k-N+1}^{k} {D}_{\text{DL},m}$ over $N$ latest control operations is larger than $\alpha D_\text{th}$ for $0<\alpha < 1$. The condition (i) implies sufficiently large HJB NN training samples, and (ii) indicates that the UAV is far away from the BS. The effectiveness of \textsf{oHJB} is validated by numerical evaluations in the next section.

	\begin{figure*}[t]
		\centering
		\includegraphics[width= \linewidth, height = 9cm, trim=2.8cm .5cm 2.8cm .5cm,clip]{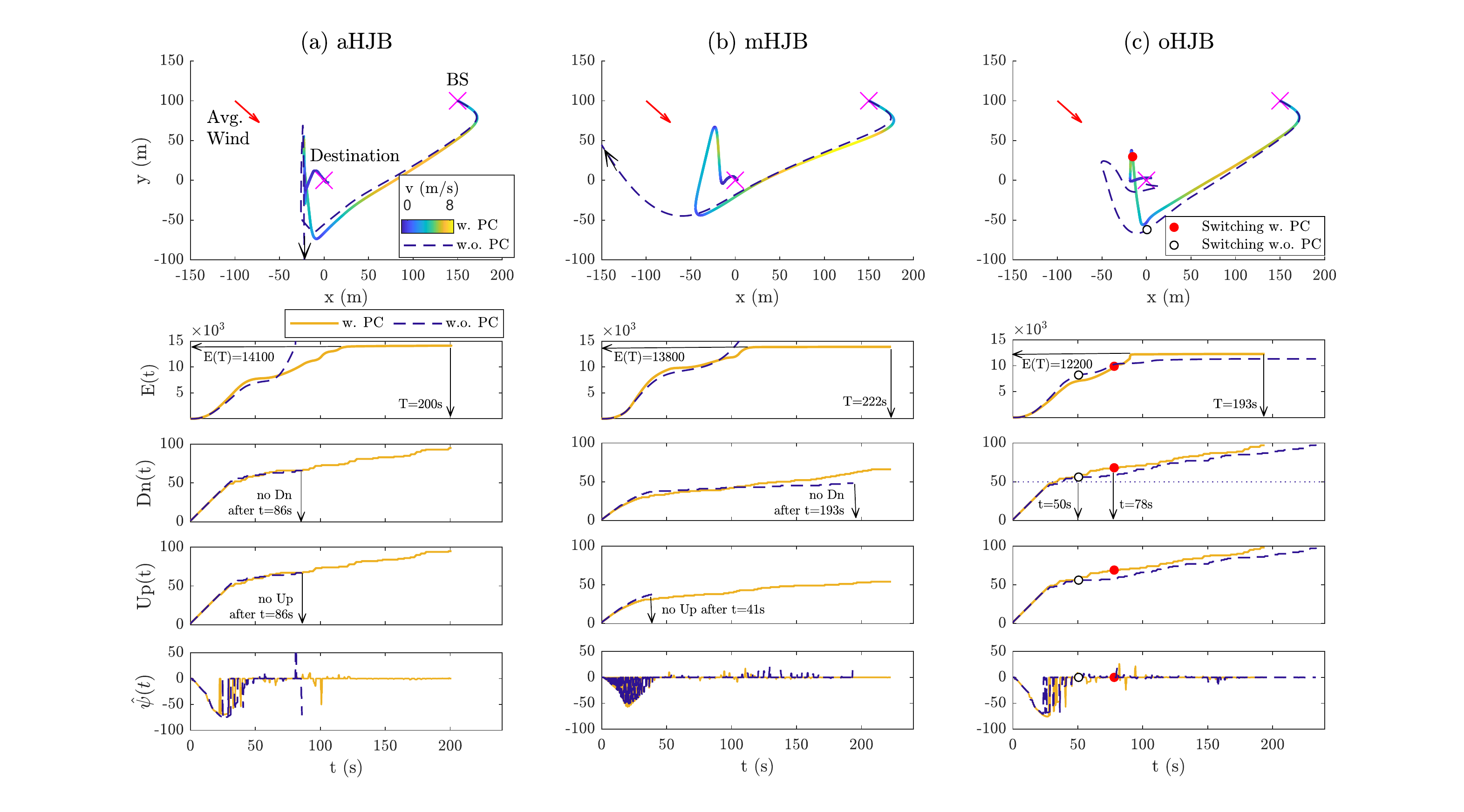} \vskip -5pt
		\caption{Comparison of (a) \textsf{aHJB}, (b) \textsf{mHJB}, and (c) \textsf{oHJB} control algorithms. Each row shows: (i) the UAV trajectories, (ii) the amount $\text{Dn}(t)$ of downloaded UAV states, (iii) the amount $\text{Up}(t)$ of uploaded control data, (iv) the output $\hat{\psi}(t)$ of the HJB NN, and (v) the motion energy $E(t)$ of the UAV, respectively. As shown by the trajectories, \textsf{oHJB} achieves the shortest travel path with the minimum motion energy $E(T)$ and the travel time $T=193$s. Initially, \textsf{oHJB} uploads optimal actions, which yields longer controllable time (i.e., larger $\text{Up}(t)$) and more training samples (i.e., larger $\text{Dn}(t)$), thereby training the HJB NN better (i.e., less fluctuating $\hat{\psi}(t)$). After the switching point (i.e., circle markers) of \textsf{oHJB}, this better trained HJB NN is uploaded to the UAV, thereby locally carrying out optimal actions even when the UAV is far away from the BS.
		}
		\label{fig:03}
	\end{figure*}

	\section{Numerical results}
	In this section, we compare the performance of the \textsf{aHJB}, \textsf{mHJB}, and \textsf{oHJB} control methods. We consider that both UAV and BS are initially located at $ (150, 100) $m, and the UAV's altitude is fixed as $h=30$m. The wind dynamics is characterized by the average wind velocity $ v_o = (1, -1) $m/s and the covariance matrix $V_o = 0.1 I $. 
	
Following \cite{b2,shiri2019massive}, the HJB model is a single hidden layer model as \eqref{Eq:07}, where each hidden node's activation function corresponds to each non-scalar term in a polynomial expansion. The polynomial is heuristically chosen as: $(1+ x_i(t) + v_{x,i}(t))^6 + (1+ y_i(t) + v_{y,i}(t))^6$ for $\sigma(s_i(t))$, where $r_i(t) = [x_i(t),y_i(t)]^{\T} $ and $ v_i(t) = [v_{x,i}(t),v_{y,i}(t)]^{\T}$, thus the model size is $ M\!=\!  54 $.

Other settings are summarized as follows: \small{$P_d= 20\text{dBm}$, $P_{\text{UL},o}= 23\text{dBm}$ and $ P_{\text{UL},\text{max}} = 26\text{dBm} $, $f_c= 2$GHz, $W_i=2$MHz, $N_i=-118$ dBm/Hz, $ b = 10\text{bytes} $, $\text{Dn}_\text{th}=50$, $c_0=0.1$, $c_1 = c_2 = 0.015$, $c_3 = 0.005 $, $ c_{\Omega} = 0.5 $, $\mu = 0.01$, $D_{\text{th}}=2\text{s}$, $\alpha=0.2$, $N=5$, {\normalsize and} $\Delta t = 0.1\text{s} $}\normalsize~for discretizing time in simulations.

\figurename{ \ref{fig:03}} compares (a) \textsf{aHJB}, (b) \textsf{mHJB}, and (c) \textsf{oHJB}. As observed by the UAV trajectories in the first row, \textsf{oHJB} achieves the shortest travel path, even without power control (w.o. PC, dashed curves) as opposed to \textsf{aHJB} and \textsf{bHJB} that fail to reach the destination. Furthermore, as shown by the second row of Fig.~2, \textsf{oHJB} also achieves up to $13.1$\% shorter travel time with $13.5$\% less motion energy $E(T)$ compared to \textsf{aHJB} and \textsf{mHJB}, where $E(t)=\int_{\tau=0}^t c_2\| v(\tau)\|^2 + c_3 \|a(\tau) \|^2d\tau$. The rationale behind these results comes from the HJB NN training and channel characteristics as detailed next.

Comparing \textsf{aHJB} and \textsf{mHJB}, in the early phase we observe that \textsf{aHJB} allows the HJB NN to feed more training samples. The reason is that uploading actions yield $27$x smaller uplink payload sizes than uploading the HJB NN models. This enables more frequent downloading of the UAV states and more trained HJB NN, as seen by higher $\text{Dn}(t)$ and less fluctuating $\hat{\psi}(t)$ in Fig.~2. 

On the other hand, it is remarkable that \textsf{mHJB} achieves comparable travel time with even lower $E(T)$, although its HJB NN is less trained than \textsf{aHJB}'s. The less accurate HJB NN output is compensated in the latter phase, in which uploading the HJB NN model enables the UAV to locally carry out its optimal control, even when the uplink connectivity is poor, i.e., low uploaded data amount $\text{Up}(t)$. Nevertheless, without PC, both \textsf{aHJB} and \textsf{bHJB} cannot reach the destination, since the uplink and downlink connection is entirely lost after a certain time period.

To enjoy both benefits of \textsf{aHJB} and \textsf{mHJB}, \textsf{oHJB} follows \textsf{aHJB} in the early phase, and switches to \textsf{mHJB} in the latter phase. These two phases are switched at the circles in Fig.~2c, at which the HJB NN has observed more than $50$ training samples and the average downlink delay during $5$ latest receptions has exceeded $ 20 $\% of $D_\text{th}$. Consequently, in the early phase, \textsf{oHJB} downloads as many training samples as \textsf{aHJB} (see $\text{Dn}(t)$), and achieves the same level of HJB NN training (see $\hat{\psi}(t)$). Utilizing this well trained HJB NN, in the latter phase, \textsf{oHJB} achieves the fastest travel time with the lowest motion energy. In sharp contrast to \textsf{aHJB} and \textsf{mHJB}, without PC, \textsf{oHJB} still makes the UAV reach the destination, with $24$\% longer travel time and similar motion energy compared to the case with PC.

\section{conclusion}	
In this letter, to remotely control a single UAV, we proposed an online NN learning based control algorithm \textsf{oHJB} that follows an action uploading based control method \textsf{aHJB} in the initial phase, followed by an NN model uploading based control scheme \textsf{mHJB}. Extending this framework to incorporating deep NN architectures \cite{Park:2018aa} with multiple UAVs could be an interesting topics for future research.

	\bibliographystyle{IEEEtran}
	

\end{document}